\documentclass[prl,twocolumn,amsmath,amssymb,floatfix,superscriptaddress]{revtex4}

\usepackage{graphicx}
\usepackage{multirow}
\newcommand{\comments}[1]{}
\begin{document}

\title{Deterministic entanglement of photons \\ in two superconducting microwave resonators}

\author{H. Wang}
\affiliation{Department of Physics, University of California, Santa Barbara, CA 93106, USA}
\affiliation{Department of Physics and Zhejiang California International NanoSystems Institute, \\ Zhejiang University, Zhejiang 310027, China}
\author{Matteo Mariantoni}
\affiliation{Department of Physics, University of California, Santa Barbara, CA 93106, USA}
\author{Radoslaw C. Bialczak}
\affiliation{Department of Physics, University of California, Santa Barbara, CA 93106, USA}
\author{M. Lenander}
\affiliation{Department of Physics, University of California, Santa Barbara, CA 93106, USA}
\author{Erik Lucero}
\affiliation{Department of Physics, University of California, Santa Barbara, CA 93106, USA}
\author{M. Neeley}
\affiliation{Department of Physics, University of California, Santa Barbara, CA 93106, USA}
\author{A. O'Connell}
\affiliation{Department of Physics, University of California, Santa Barbara, CA 93106, USA}
\author{D. Sank}
\affiliation{Department of Physics, University of California, Santa Barbara, CA 93106, USA}
\author{M. Weides}
\affiliation{Department of Physics, University of California, Santa Barbara, CA 93106, USA}
\author{J. Wenner}
\affiliation{Department of Physics, University of California, Santa Barbara, CA 93106, USA}
\author{T. Yamamoto}
\affiliation{Department of Physics, University of California, Santa Barbara, CA 93106, USA}
\affiliation{Green Innovation Research Laboratories, NEC Corporation, Tsukuba, Ibaraki 305-8501, Japan}
\author{Y. Yin}
\affiliation{Department of Physics, University of California, Santa Barbara, CA 93106, USA}
\author{J. Zhao}
\affiliation{Department of Physics, University of California, Santa Barbara, CA 93106, USA}
\author{John M. Martinis}
\affiliation{Department of Physics, University of California, Santa Barbara, CA 93106, USA}
\author{A. N. Cleland}
\email[To whom correspondence should be addressed. E-mail: ]{anc@physics.ucsb.edu}
\affiliation{Department of Physics, University of California, Santa Barbara, CA 93106, USA}

\date{\today}

\begin{abstract}
Quantum entanglement, one of the defining features of quantum mechanics, has been demonstrated in a variety of nonlinear spin-like systems.
Quantum entanglement in linear systems has proven significantly more challenging, as the intrinsic energy level degeneracy associated with linearity makes quantum control more difficult. Here we demonstrate the quantum entanglement of photon states in two independent linear microwave resonators, creating $N$-photon NOON states as a benchmark demonstration. We use a superconducting quantum circuit that includes Josephson qubits to control and measure the two resonators, and we completely characterize the entangled states with bipartite Wigner tomography. These results demonstrate a significant advance in the quantum control of linear resonators in superconducting circuits.
\end{abstract}

\maketitle

Quantum superposition and entanglement have been demonstrated experimentally using spin-like physical systems ranging from atoms to electronic circuits \cite{haroche:2006, blatt:2008, hanson:2007, neumann:2008, clarke:2008, neeley:2010, dicarlo:2010}. These systems all display strong nonlinearity, and are used because this nonlinearity allows straightforward quantum control by classical means. The quantum control of linear systems, exemplified by the harmonic oscillator, is by contrast more difficult, and has only been achieved using nonlinear intermediaries: Atoms \cite{haroche:2006, mabuchi:2002} to control optical cavities, ions to control ion motion \cite{leibfried:2003, jost:2009}, and superconducting qubits to control photons in microwave resonators \cite{schoelkopf:2008, hofheinz:2008, hofheinz:2009, wang:2009b}. Quantum entanglement of cavity photons still presents a significant challenge: Experiments have demonstrated maximally-entangled photons in different polarization modes of the same cavity \cite{rauschenbeutel:2001, papp:2009}, but the entanglement of photons in two physically distinct cavities \cite{mariantoni:2008, merkel:2010, strauch:2010} has proven more elusive.

\begin{figure}
\includegraphics[clip=True,width=0.43\textwidth]{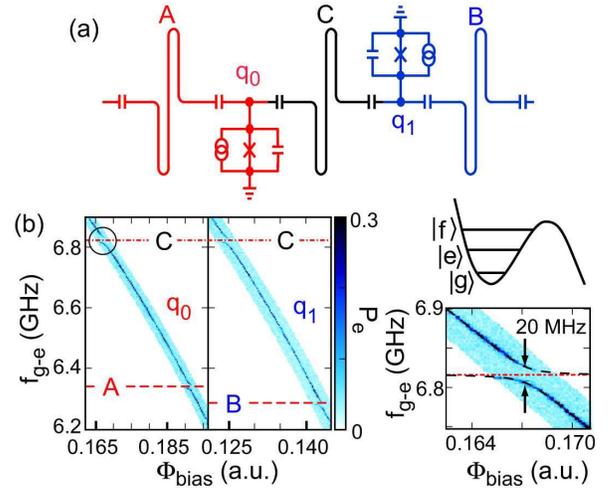}
\caption{\label{fig.device}(Color online) (a) Device circuit schematic. Coupling resonator $C$ is connected to $q_0$ and $q_1$
using 20 MHz coupling capacitors, while storage resonator $A$ ($B$) is
connected to $q_0$ ($q_1$) through a 17 MHz coupling capacitor.
(b) Qubit spectroscopy, showing probability $P_e$ (color bar) vs. microwave frequency and flux bias for each
qubit. Avoided-level crossings near $6.8$\,GHz (dash-dot
lines) are due to the coupling resonator
$C$ and near 6.3 GHz (dashed lines) due to each qubit's storage resonator. Lower right panel shows magnified view of circled area, upper right panel shows three qubit levels.}
\end{figure}

Here we show the deterministic generation of entangled photon states in two spatially-separated microwave resonators, achieved by manipulating the photon states with a pair of superconducting phase qubits. We use as a benchmark the generation of NOON states \cite{boto:2000, nagata:2007, shalm:2009, jones:2009, afek:2010}, comprising a total of $N$ photons in the two resonators ($A$ and $B$), entangled in the quantum state
\begin{equation}
\left | \psi \right \rangle = \frac{1}{\sqrt{2}} \left (
\left | N \right \rangle_A \left | 0 \right \rangle_B +
\left | 0 \right \rangle_A \left | N \right \rangle_B
\right ) ,
 \label{eq.1}
\end{equation}
with $N$ photons in resonator $A$ and zero in $B$, superposed with the state with the occupation numbers reversed. Such a state has the same degree of entanglement as the Bell
state, but with $N$ excitations. We also generate MOON states, in which, e.g., resonator $A$ has $M$ or zero quanta, entangled with resonator $B$ with zero or $N$ quanta.We fully characterize the two-resonator photon states using bipartite Wigner tomography, which represents a non-trivial extension of single-cavity Wigner tomography \cite{haroche:2006, leibfried:2003, hofheinz:2008, hofheinz:2009, wang:2009b}, and allows us to distinguish entanglement from an incoherent ensemble.

To accomplish this goal, we developed a new quantum circuit comprising two superconducting phase qubits \cite{martinis:2009} and three microwave resonators. A sketch of the circuit topology is shown in Figs. \ref{fig.device}(a). The circuit includes a coupling resonator $C$, connected to both qubits, and two state storage resonators $A$ and $B$, each coupled to one qubit. The resonator frequencies are all different which allows us to frequency-select the qubit-resonator interactions.

\begin{figure}
\includegraphics[clip=True,width=0.43\textwidth]{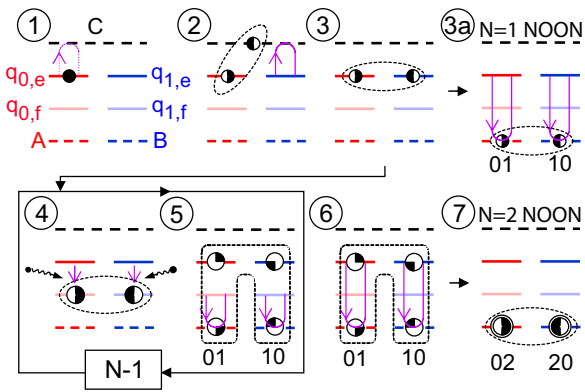}
\caption{\label{fig.stateprep}(Color online) NOON state preparation sequence. Resonators are
represented by dashed lines, and the qubit $| g
\rangle \leftrightarrow | e \rangle$ and $| e
\rangle \leftrightarrow | f \rangle$ transitions by
dark and light color solid lines, respectively. (1):
$q_0$ is excited to $| e \rangle$ and
half-swapped to $C$, generating the Bell state $| e 0 \rangle + | g 1 \rangle$. (2): Coupling resonator
swapped to $q_1$, generating $| eg \rangle + | ge \rangle$. (3a): $N = 1$ NOON
state $| 1 0 \rangle + | 0 1\rangle$ generated by fully swapping each qubit to its storage resonator. For higher $N$ states: (4): Qubits
excited to $| fg \rangle + | gf\rangle$. (5): One photon swapped into storage resonators, generating
$| eg 1 0 \rangle + | ge 0 1 \rangle$. Steps (4) and (5) are repeated $N-1$ times, generating $|eg( N -
1 ) 0 \rangle + | g e 0 ( N - 1 ) \rangle$. (6): Final
photon transfer generates (7) $N = 2$ (or higher $N$)
NOON state.}
\end{figure}

The basic method for generating two-resonator entangled states, illustrated in Fig. \ref{fig.stateprep}, is to excite and then entangle the two qubits using the coupling resonator. We can swap the resulting Bell state $|eg\rangle +|ge\rangle$ to the two storage resonators, creating an $N=1$ NOON state $\left | 1 0 \right \rangle + \left | 0 1 \right \rangle$. If we want to generate higher $N$ photon states, we instead selectively excite each qubit to its next higher energy level $\left |  f \right \rangle$ (Fig. \ref{fig.device}(b)), generating the state $\left | fg \right \rangle + \left | gf \right \rangle$, thus using the qubits as ``qutrits'' \cite{you:2007, neeley:2009}. The required microwave excitation is selective, due to the anharmonicity of the qubits. The qubit excitation is then swapped to each storage resonator through the qubit $|f\rangle \leftrightarrow |e\rangle$ transition, creating a four-fold entangled state $\left | eg 1 0 \right \rangle + \left | ge 0 1 \right \rangle$, where the first two letters indicate the qubit
states, and the second two numbers the storage resonator states. We then re-excite the qubits to their $|f\rangle$ states, and again swap the excitation to the resonators,
generating $\left | eg 2 0 \right \rangle + \left | ge 0 2 \right \rangle$. This process can be repeated until the entangled state has $N-1$ photons. In the final step, each qubit's $|e\rangle \leftrightarrow |g\rangle$ transition is brought on resonance with the corresponding storage resonator, swapping the last excitation and leaving the system in $\left | gg N 0 \right \rangle + \left | gg 0 N \right \rangle = |gg\rangle \otimes (|N0\rangle+|0N\rangle)$, an $N$-photon NOON state.

We analyze the final
resonator state using the qubits as probes. The simplest
analysis uses a coincidence measurement: We bring the
qubits into resonance with their corresponding storage
resonators for an interaction time $\tau$, after which both
qubits are measured simultaneously. The preparation and
measurement sequence is repeated $\sim 10^3$ times,
yielding the joint-qubit state probabilities, $P_{g g}$,
$P_{ge}$, $P_{e g}$, and $P_{e e}$, where $P_{g e}$ is the probability of
measuring the first qubit in its ground state with the second
qubit in its excited state, and so on. We then vary the
interaction time $\tau$, capturing the evolution of these
probabilities. If a resonator has $n$ photons, the $n$th photon
will swap between the qubit and resonator at a
rate scaling as $\sqrt{n}$, while for more
complicated states, the interaction is a sum of components
oscillating at their respective frequencies, weighted by the photon occupation probabilities \cite{hofheinz:2009,wang:2009b}.

\begin{figure}
\includegraphics[clip=True,width=0.43\textwidth]{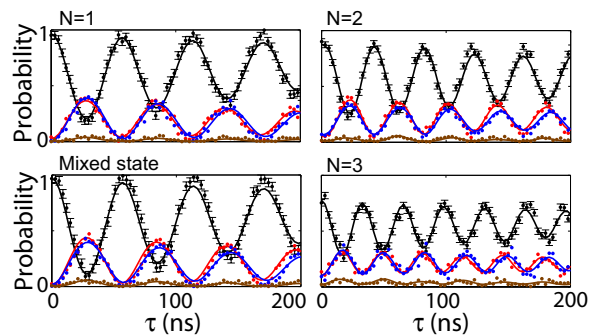}
\caption{\label{fig.statecoinc}(Color online) Qubit coincidence probability measurements for $N = 1$, $2$, and $3$ NOON
states, and for a mixed state. $P_{g e}$ (blue) and $P_{ e g}$ (red)
oscillate with interaction time $\tau$ at a rate $\propto \sqrt{N}$, while
$P_{e e}$ (brown) always remains small. For the mixed state, behavior is nearly identical to the $N = 1$
NOON state. Lines are fits to the data. Statistical
errors, from the measured probability spread of $\sim 2-3
\%$, are shown only for $P_{g g}$ \cite{PRLsupplementary}. Horizontal and vertical axes are same for all plots.}
\end{figure}

For resonators entangled in a NOON state, a joint measurement
should correspond to either $N$ photons in one resonator and
zero in the other, or to the reverse situation; the measurement
of the qubits ``collapses'' the system onto one or the other
outcome. Thus in one measurement at most
one of the qubits will be in the excited state. When
averaged over many measurements, the maximum probability
of measuring a particular qubit in $\left | e \right
\rangle$ is $1 / 2$, while the probability of measuring both
qubits in $\left | e \right \rangle$ should be zero. Therefore we expect that $P_{ e g}$
and $P_{g e}$ will oscillate between $0$ and $1/2$, $P_{e e}$ will be zero, and $P_{g g}$ should equal $1-P_{ge}-P_{eg}-P_{ee}$. 

\begin{figure*}
\includegraphics[clip=True,width=0.86\textwidth]{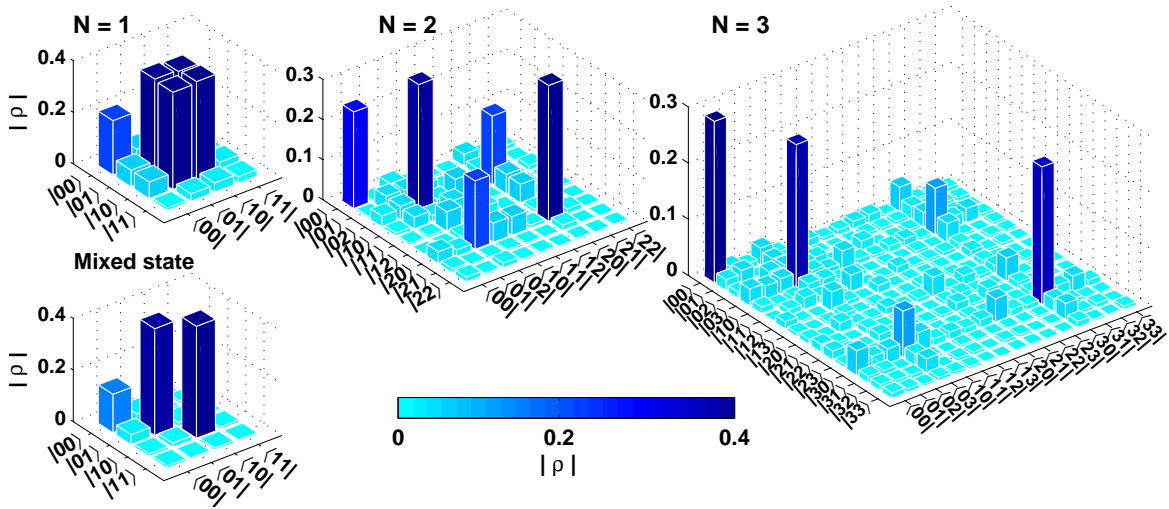}
\caption{\label{fig.stateTomo} (Color online) NOON and mixed state density matrix amplitudes, reconstructed in the photon-number basis
from bipartite Wigner tomography; states are labeled $| mn \rangle$
where $m$ is the photon number in resonator $A$ and $n$ that in $B$. Bar heights and colors
represent matrix element amplitudes. The dominant amplitudes for all three states are in the
expected locations, although the off-diagonal elements decrease with $N$, due to the finite qubit
dephasing time. Errors for the density matrix elements are not
shown but are small \cite{PRLsupplementary}.}
\end{figure*}

Coincidence measurements are shown in Fig. \ref{fig.statecoinc}
for NOON states up to $N = 3$, and are consistent with these
expectations. The oscillations seen in these measurements are however insufficient proof of resonator
entanglement, as an incoherent mixed state can give the same
results. To demonstrate this, we have controllably generated a
synthetic ensemble comprising a $50 \, \%$ population of $| 10
\rangle$ states and $50 \, \% \, | 0 1 \rangle$ states \cite{PRLsupplementary}. Coincidence measurements
(Fig. \ref{fig.statecoinc}) of this synthetic mixed state
generate outcomes identical to those of the $N = 1$ NOON
state.

A more complete resonator measurement, that can resolve entangled from mixed states, uses bipartite Wigner
tomography, a nontrivial extension of single resonator
tomography \cite{leibfried:2003,hofheinz:2009, wang:2009b}.
This involves injecting a coherent Gaussian microwave pulse into
each of the storage resonators, with controlled amplitude and phase,
displacing the resonator states in phase space. The resonators
are then simultaneously measured with a joint probability
measurement, now as a function of the amplitude and phase of
the coherent pulses. From the complete set of measurements, the
two-resonator density matrix can be calculated \cite{PRLsupplementary}.

In Fig. \ref{fig.stateTomo} we display the amplitudes of the density matrices
measured for resonator NOON states up to $N = 3$, as well as for the mixed state. While there
are non-idealities, the desired non-zero matrix elements are
clearly apparent for the NOON states, while for the mixed state, the density matrix has
only zero-valued off-diagonal elements. The state preparation fidelities, $F =
\langle \psi | \rho | \psi \rangle$, are found to be $0.76 \pm 0.02$ ($N = 1$), $0.50 \pm 0.02$ ($N =
2$), and $0.33 \pm 0.02$ ($N = 3$). For $N = 1$, the most probable entanglement of
formation \cite{audenaert:2001} is $EOF = 0.51$, while for $N=2$ and $N=3$, $EOF = 0.31$ and 0.28, respectively; for the mixed state, the $EOF$ is zero. We also calculate the negativity
$N_e ( \rho ) = \sum_j { \max { ( 0 ,
\mu_j ) } }$, where $\mu_j$ are the eigenvalues of the
partial transpose $\rho^{\rm PT}$ of the density
matrix \cite{shalm:2009, peres:1996}, and $N_e ( \rho ) > 0$ indicates entanglement. The negativities are found to be $0.56 \pm 0.03$ ($N = 1$), $0.32
\pm 0.03$ ($N = 2$), and $0.27 \pm 0.01$ ($N = 3$); for the mixed state we find zero with an upper bound of 0.001. The decrease of these values with photon number $N$ is compatible with expectations: The state
preparation requires phase-coherence of the four-element
entangled states for most of the preparation sequence, which is limited by the qubit coherence time 
$T_2$ \cite{PRLsupplementary}. Other
than this technical limitation, the deterministic
generation is completely scalable
to large $N$.

\begin{figure}
\includegraphics[clip=True,width=0.43\textwidth]{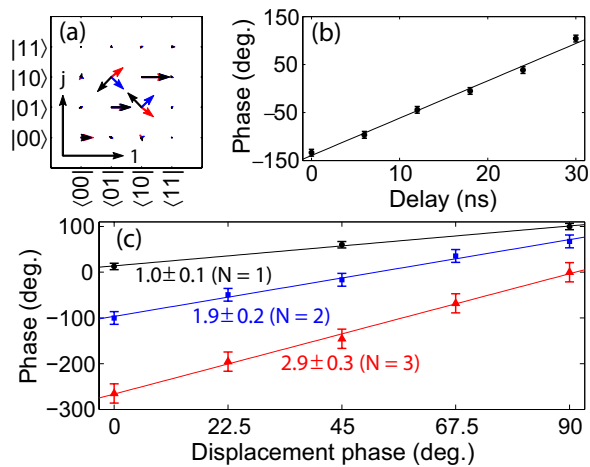}
\caption{\label{fig.phase} (Color online) Phase sensitivity of NOON
states up to $N = 3$. (a) Density matrix for $N = 1$ NOON state at different times after qubit entanglement. Each element is represented by an arrow, with orientation determined by the phase angle (scale on bottom left). Off-diagonal elements rotate with delay time (black: $0$\,ns, blue: $12$\,ns, red: $24$\,ns), due to frequency difference between the qubit operating point and resonator. Dephasing causes decrease in amplitude of off-diagonal elements. (b) Phase angle of the upper left non-zero off-diagonal element in (a) versus delay time. Line is fit to a rotation rate of $\Delta f = 21.6 \pm 0.8$\,MHz, which corresponds well to set frequency difference. Error bars indicate maximum phase uncertainty. (c) Rotation angle of off-diagonal element versus controlled, additional phase angle used in coherent state displacement. Lines are fits (slopes indicated on plot), consistent with expected phase
sensitivity. Confidence bounds are as in (b), with uncertainty increasing with $N$ due to increased phase sensitivity.}
\end{figure}

A hallmark of NOON states is their rapid phase
evolution \cite{boto:2000, nagata:2007, jones:2009, afek:2010},
which can be verified by Wigner tomography using two distinct methods. For the $N = 1$
state, after entangling the qubits in a Bell state, we wait
for a variable time and then swap the state into the storage
resonators. The density matrices measured at three different
delay times are shown in Fig. \ref{fig.phase}(a). The phases of
the off-diagonal elements rotate with time due to the
qubit-resonator frequency difference, as in
Fig. \ref{fig.phase}(b), showing the expected linear dependence.

This phase-measurement method suffers from the short qubit
dephasing time. A second method is to change the phase
reference for the coherent pulses used in the Wigner
tomography, avoiding storage of the state in the qubit. We add
an additional phase to the pulses applied to
resonator $A$ only. The
resulting density matrices show the expected rotation of the
off-diagonal elements. In Fig. \ref{fig.phase}(c) we plot the
off-diagonal phase angle for different $N$; the $N = 3$
state evolves three times faster than the $N = 1$ state, as expected.

We also used tomography to measure the NOON state decay \cite{PRLsupplementary}. We find that the off-diagonal
elements decrease at approximately the same rate as the
diagonal elements, with a decay time $\tau_D \approx {}
3\,\mu$s, consistent with a Markovian
environment \cite{wang:2009b}.

We note that the generation sequence allows an additional flexibility: We can add
different numbers of quanta to the resonators, thus generating MOON
states, $| \psi \rangle = | M 0 \rangle + | 0 N \rangle$.
An example with $M = 2$ and $N = 1$ is detailed in \cite{PRLsupplementary}.

The capabilities we have demonstrated here, generating complex
entangled photon states in two resonators, hold promise for
new quantum architectures in which superconducting resonators
play a more central role. The protocol can be extended in a
natural way to entangle larger numbers of resonators, allowing, e.g., the direct generation of resonator GHZ and W states \cite{leibfried:2003, blatt:2008, nagata:2007, neumann:2008, papp:2009, jones:2009}. The longer coherence times achievable in superconducting resonators will be of direct utility in performing more
complex quantum algorithms, furthering the capabilities of superconducting quantum circuit
architectures.

\textbf{Acknowledgments.} We thank K. Audenaert for assistance in entanglement-of-formation calculations. This work was supported by IARPA
under ARO award W911NF-04-1-0204. M.M.~acknowledges support
from an Elings Postdoctoral Fellowship. H.W. acknowledges partial support by the Fundamental Research Funds for the Central Universities in China (Program No. 1A3000*172210301[26]). Devices were made at the UC Santa Barbara Nanofabrication Facility, a part of the NSF-funded National Nanotechnology Infrastructure Network.

\clearpage
\renewcommand\thefigure{S\arabic{figure}}
\renewcommand\theequation{S\arabic{equation}}
\renewcommand\thetable{S\arabic{table}}

\setcounter{figure}{0}
\setcounter{equation}{0}
\setcounter{page}{1}
\setcounter{table}{0}

\begin{center}
{\noindent {\bf Supplementary Material for\\
``Deterministic entanglement of photons in two superconducting microwave resonators''}}
\end{center}
\section{Materials and Methods}

\renewcommand\thepage {S\arabic{page}}
The device fabrication is similar to that published
previously \cite{s_wang:2008}. The half-wavelength superconducting
coplanar waveguide resonators are made of rhenium deposited on
a $c$-axis single-crystal sapphire substrate, with a $5$
$\mu$m-wide center signal trace and 10 $\mu$m gaps to the
ground plane metallization on either side of the center trace.
We place a single lithographed shorting strap connecting the
two ground planes at the midpoint of each resonator to improve the quality of the grounding. This point
is a voltage node for the fundamental half-wave resonant mode, so that there is minimal additional dielectric loss from the shorting strap's underlying
amorphous Si insulating film \cite{s_oconnell:2008}.

In the circuit layout, the coupling resonator $C$ is designed
to have a higher resonance frequency than the two state storage
resonators $A$ and $B$. This prevents the
qubit frequencies from having to cross the $C$ resonator frequency during
NOON state amplification. The two storage resonators $A$ and
$B$ are designed with slightly different resonance frequencies,
to avoid possible interference between the resonators. The full
frequency span in the design was chosen to be about $550$ MHz,
within the dynamic range of our custom microwave electronics. The
two superconducting phase qubits and coplanar waveguide
resonators are fabricated together, using our standard
multi-layer process \cite{s_martinis:2009}. We use interdigitated
coupling capacitors between the qubits and the resonators,
calculated to each have a coupling capacitance of 1.9 fF. The
actual coupling strengths vary slightly with resonator
frequency; the detailed component parameters are listed in
Table \ref{tab.para}.

\begin{table*}[ht]
\begin{tabular}{|lccccc|}
\hline \hline
circuit & $f_{\rm r}$ & $f_{\left| g \right\rangle \leftrightarrow \left| e \right\rangle}$ & $f_{\rm nonlinear}$ & $T_1$ & $T_{\phi}$\\
component & (GHz) & (GHz) & (GHz) & (ns) & (ns)\\\hline
$A$ & $6.340$ & - & - & $3500$ & $\gg T_1$\\
$B$ & $6.286$ & - & - & $3300$ & $\gg {} T_1$\\
$C$   & $6.816$ & - & - & $3400$ & $\gg {} T_1$\\ \hline
$q_0$ & - & $\sim$ $6.65$ & $\approx$ $0.20$ & $450$ & $200 - 300$\\
$q_1$ & - & $\sim$ $6.58$ or $6.68$ & $\approx$ $0.20$ & $320$ & $200 - 300$ \\ \hline \hline
\multicolumn{2}{|l}{Coupling strength} & $q_0 \leftrightarrow A$ & $q_0 \leftrightarrow C$ & $q_1 \leftrightarrow B$ & $q_1 \leftrightarrow C$\\
\multicolumn{2}{|l}{$g/\pi$ (MHz)} & $17.8$ & $20.0$ & $17.4$ & $20.0$\\ \hline \hline
\end{tabular}
\caption{\label{tab.para} Resonator and qubit parameters. The qubit parameters are quoted for when the qubits are off-resonance (see below). The qubit nonlinearity $f_{\rm nonlinear}$ is the
frequency difference between the $| g \rangle \leftrightarrow | e \rangle$ and $| e \rangle \leftrightarrow | f \rangle$ transitions. The phase coherence time $T_{\phi}$ is obtained using a
Ramsey interference experiment, which yields the Ramsey time $T_2$, from which we calculate $1 / T_{\phi} = 1 / T_2 - 1 / ( 2 T_1 )$. $T_\phi$ measured for resonators similar to those used here \cite{s_wang:2009b} satisfies $T_{\phi} \gg T_1$; we assume the same applies here. $T_{\phi}$ for the qubits decreases with increasing length of the pulse sequence due to the $1 / f$ nature of the phase noise. For most state generation sequences used in this experiment, of typical length $50$ to $100$ ns, the qubit $T_{\phi}$ is in the range of $200$ to $300$ ns. Coupling strengths correspond to the measured splitting in frequency units, with $g$ appearing in the Hamiltonian shown in Eq.~(\ref{eq.ham}).}
\end{table*}

\section{Generation Sequence Tune-Up}

\begin{figure}[t]
\includegraphics[clip=True,width=0.4\textwidth]{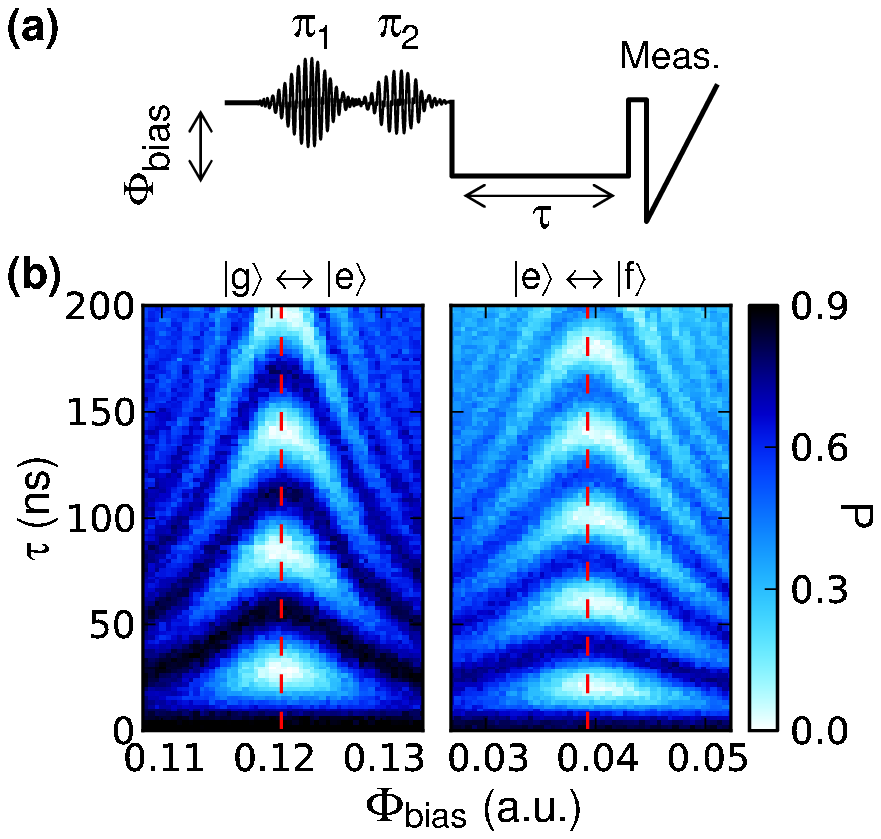}
\caption{\label{fig.swap} (Color online) One-photon swaps between qubit
$q_0$ and resonator $A$. (a) Pulse sequence for
measuring swaps between the qubit $| e \rangle \leftrightarrow | f \rangle$ transition and resonator $A$. When off-resonance, the
qubit is excited by a pair of $\pi$ pulses, with the first
pulse $\pi_1$ taking the qubit from $| g \rangle$ to $| e
\rangle$ and the second pulse $\pi_2$ making the transition
from $| e \rangle$ to $| f \rangle$. The qubit $| e
\rangle \leftrightarrow | f \rangle$ transition is
then tuned close to resonance with resonator $A$ and the qubit
and resonator left to interact for a time $\tau$. The triangular pulse
at the end is used to measure the probability of the qubit
being in the $| f \rangle$ state at the end of the
sequence. (b) Left: Qubit $| g \rangle \leftrightarrow | e \rangle$ swaps with resonator
$A$. The qubit is prepared as in (a), but without the
second pulse $\pi_2$. The qubit $| e  \rangle$ probability (color
bar) is plotted as a function of interaction time $\tau$ and
frequency tuning $\Phi_{\rm bias}$. Oscillations in
probability are due to the qubit excitation swapping with the
resonator. Right: Qubit $| e \rangle \leftrightarrow |
f \rangle$ swaps with resonator $A$; the preparation of
the qubit is as in (a) and the plot shows the $| f \rangle$-state probability as a function of interaction time
$\tau$ and qubit flux bias $\Phi_{\rm bias}$. Red dashed lines
indicate the on-resonance $\Phi_{\rm bias}$ values used in
the experiment. The swap frequency for the on-resonance $| e
\rangle \leftrightarrow | f \rangle$ transition is
$1.403 {} \approx {} \sqrt{2}$ times that for the on-resonance
$| g \rangle \leftrightarrow | e \rangle$ transition.}
\end{figure}

The time required for each qubit-resonator $i$-SWAP is
calibrated separately. The swap times obtained from these
calibrations scale correctly as $\sqrt{n}$ with the number of
photons $n$ in the resonator \cite{s_hofheinz:2009} and also depend on
the state of the qubit. Examples of the swap calibrations for a
one-photon swap between qubit $q_0$ and resonator $A$ are
shown in Fig.~\ref{fig.swap}, for both the  $| g
\rangle \leftrightarrow | e \rangle$ and $| e
\rangle \leftrightarrow | f \rangle$ transitions.
The swap time for the $| e \rangle \leftrightarrow | f \rangle$ transition is approximately $1 / \sqrt{2}$ times
that for the $| g \rangle \leftrightarrow | e \rangle$
transition. This scaling is as expected, as the multi-level
phase qubit can be well-approximated as a weakly nonlinear
harmonic oscillator for the energy levels confined by the qubit's
metastable potential well. The scaling confirms that we can use
harmonic-oscillator-like raising and lowering operators for the
three-level qubit, as was done in the numerical simulations (see
below).

\begin{figure*}[t]
\includegraphics[width=0.9\textwidth]{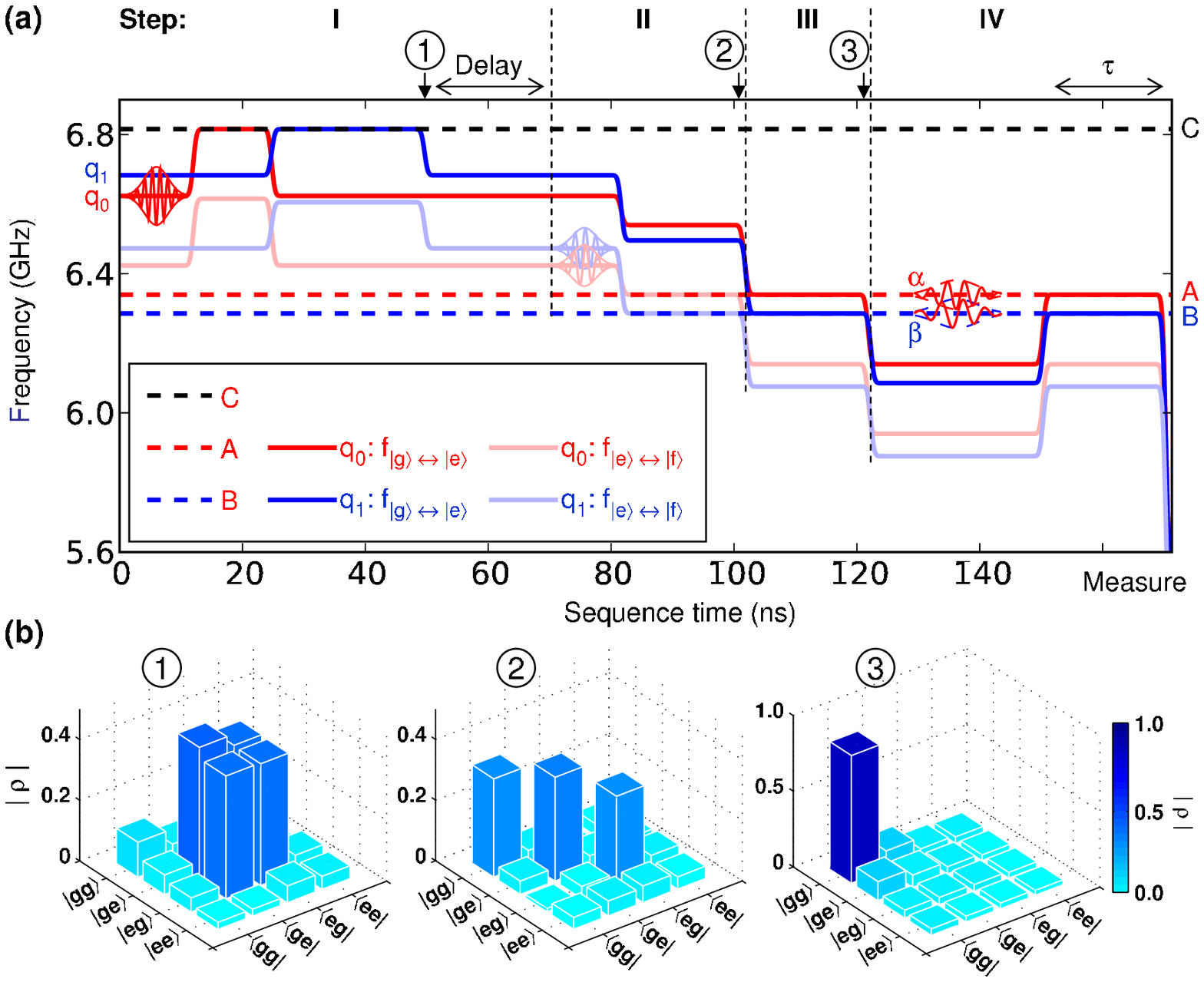}
\caption{\label{fig.qubitTomo}(Color online) NOON state preparation. (a) Sequence
to generate and measure $N = 2$ NOON state. Qubit
transition frequencies $| g \rangle \leftrightarrow | e \rangle$ (dark-color solid lines) and $| e
\rangle \leftrightarrow | f \rangle$ (light solid
lines) are tuned to control coupling of each
qubit to resonators (dashed lines); $q_0$ in red, $q_1$ in blue. Vertical lines divide sequence into steps; circled numbers indicate points where qubit tomography was performed. Step {\bf
I}: A $\pi_{| g \rangle \rightarrow | e \rangle}$ pulse
excites $q_0$ to $\left| e \right\rangle$, followed
by a $\sqrt{i\textrm{-}\mathrm{SWAP}}$ entangling $q_0$
and resonator $C$. A full $i$-SWAP between $C$ and $q_1$
disentangles the resonator and generates the qubit Bell state
$| e g \rangle + | g e \rangle$. For the $N = 1$ NOON
state this entanglement is transferred to the storage
resonators (not shown). For $N > 1$, $\pi_{| e \rangle
\rightarrow | f \rangle}$ pulses excite each qubit to yield $| f g \rangle + | g f \rangle$. Both qubits'
$| e \rangle \leftrightarrow | f \rangle$ transitions are then
tuned to the respective storage resonators,
and an $i$-SWAP transfers one photon to
the resonators. For $N = 3$, this is repeated to
transfer a second photon (not
shown). In step {\bf III}, both qubits' $| g
\rangle \leftrightarrow | e \rangle$ transitions are
tuned to the resonators and an $i$-SWAP
transfers the remaining excitation to the
resonators, completing the generation, leaving
both qubits in their ground states. Step {\bf IV}: State analysis. The storage resonator $A$ ($B$) is displaced in phase space by a Gaussian
pulse $|\alpha\rangle$ ($|\beta\rangle$), followed by an on-resonance
interaction with each qubit's $| g \rangle \leftrightarrow | e
\rangle$ transition for a time $\tau$, followed by joint-qubit
state readout. Many repetitions of this sequence are combined to generate joint-qubit state probabilities $P_{gg}$, $P_{ge}$,
$P_{eg}$, and $P_{ee}$ versus $\tau$. The joint-qubit state
probabilities can be used to obtain the two-resonator density matrix (see text). (b) Qubit density matrices extracted
from tomography at circled points in (a). (1) Qubit Bell state. (2) Qubit state after transferring first photon into resonators. This is a state with qubits and resonators entangled, so
the density matrix is as expected. (3) Qubit state after $N = 2$ NOON state generation. Both qubits are
very nearly in their ground states, and thus can be used to measure the resonators.}
\end{figure*}

Figure~\ref{fig.qubitTomo} shows the detailed pulse sequence used
to generate and measure the $N = 2$ NOON state. Sequence steps in general are calibrated and checked separately to maximize preparation fidelity, when possible. For example, we first
optimize the qubit Bell state preparation at the end of step I in the preparation sequence detailed in Fig. \ref{fig.qubitTomo}.
The fidelity for the Bell state is above $0.80$, with entanglement of formation 0.59, which agrees well with numerical
simulations, performed using a pure dephasing time $T_{\phi}
= 300$ ns for the qubits (see below). State tomography of the qubits is also as expected: Fig.~\ref{fig.qubitTomo}(b) shows the density matrices extracted
from coupled qubit tomography, measured at different times
during the $N = 2$ NOON state preparation. We note that
at the end of the sequence, both qubits should return to their
ground states. Experimentally we observe small populations in
the excited states $| e \rangle$ due to decoherence and pulse
imperfections. The exact qubit state after the NOON state
generation is measured and is used as the initial state for the
qubits when performing Wigner tomography on the storage
resonators (see below).

\section{Numerical Simulations}

Numerical simulations were performed using the model
Hamiltonian
\begin{eqnarray}
\label{eq.ham}
H & = & \sum_{q_i = q_0,q_1}{H_{q_i}} + \sum_{j = A , B , C}{\hbar \omega_{j}\left(b_{j}^+ b_{j} + \frac{1}{2} \right)} \nonumber \\
  &   & + \sum_{(q_i,j)} {\hbar g_{q_i \leftrightarrow j} \left( a_{q_i}^+ b_{j} + a_{q_i} b_{j}^+ \right) } \nonumber \\
  &   & + \sum_{q_i = q_0 , q_1} { \frac{1}{2} \hbar \left[ \Omega_{q_i} ( t ) a_{q_i}^+ + \Omega_{q_i} ( t )^* a_{q_i} \right] },
\end{eqnarray}
where $H_{q_i}$ is the Hamiltonian of the qubit
$q_i$, $a_{q_i}^+$ and $a_{q_i}$ ($b_{j}^+$ and $b_{j}$)
are the raising and lowering operators for the 3-level qubit
$q_i$ (resonator $j$), $g_{q_i \leftrightarrow j}$ is
the coupling strength between qubit $q_i$ and resonator $j$,
with a sum over all possible qubit-resonator combinations, such
that $(q_i , j) \in \left\{ ( q_0 , A ) , ( q_0 , C ), ( q_1 , B ) , ( q_1 , C ) \right\}$ and
$\Omega_{q_i} ( t )$ is the time-dependent, two-tone
($f_{| g \rangle \leftrightarrow | e \rangle}$ and $f_{|e \rangle \leftrightarrow | f \rangle}$) microwave drive on
qubit $q_i$.

The 3-level qubit Hamiltonian $H_{q_i}$ was approximated as
\begin{equation}
H_{q_i} =
\left[\begin{array}{ccc}
0 & 0 & 0 \\
0 & h f_{| g \rangle \leftrightarrow | e \rangle} & 0 \\
0 & 0 & h f_{| g \rangle \leftrightarrow | f \rangle}
\end{array}\right]_{q_i},
\end{equation}
where for simplicity we assumed a constant nonlinearity $f_{| g \rangle \leftrightarrow | e \rangle} - f_{| e \rangle \leftrightarrow | f \rangle} = 200$ MHz, so
that $f_{| g \rangle \leftrightarrow | f \rangle} \simeq
2 f_{| g \rangle \leftrightarrow | e \rangle} - 200$
MHz. We approximated the multi-level qubit $a_{q_i}^+$ and
$a_{q_i}$ by the raising and lower operators for the
lowest three levels of a harmonic oscillator, as discussed
above.

Decoherence was approximated using the Lindblad master equation
taking into account the Markovian environment \cite{s_lindblad:1976},
where two characteristic decay times, the energy relaxation
time $T_1$ and the pure dephasing time $T_\phi$, were used
for each resonator and qubit.

The simulations do not directly account for the non-Markovian
$1 / f$ character of the phase noise in the qubits. To account
for this, we used a sequence-time dependent $T_\phi$ for each
qubit, as obtained from Ramsey interference measurements. We
used $T_\phi = 300$ ns for $\sim$50 ns-long
sequences and $T_\phi = 200$ ns for $\sim$100
ns-long sequences. The resulting simulations agree reasonably
well with the experimental measurements.

\section{Bipartite Wigner Tomography}

\subsection{Displacement Pulses}
The bipartite Wigner tomography is an extension of a method described
elsewhere \cite{s_leibfried:1996,s_hofheinz:2009,s_wang:2009b}. We displace each
resonator with Gaussian pulses $|\alpha\rangle$ and $| \beta
\rangle$ (resonator $A$ and $B$, respectively), with variable
phase and amplitude. The values of $\alpha$ and $\beta$ are
distributed over several concentric circles in the complex
plane, centered on the origin, where the distribution of values
varies approximately with the size $N$ of the NOON state. The
radii of the circles run through the set $r_j \in \{0 , 0.2
, 0.7 , 0.9 , 1.3\}$, in square-root of photon number
units \cite{s_hofheinz:2009}. The pulse values are evenly distributed
on each circle, with complex values $r_j \exp( i 2\pi \ell /
N_{r_j} )$, where $\ell$ ranges from 1 to $N_{r_j}$
and $N_{r_j}$ is an integer ranging from 1 ($r_j = 0$), 5 or 6 ($r_j = 0.2$), up to $9$ to $15$
($r_j = 1.3$), depending on the number of photons in
the NOON state.

We use every possible combination of values of $\alpha$ and
$\beta$ distributed over the circles of the same radius for
tomography, i.e., for each value of $\alpha$, we use all $\beta$
values with the same amplitude as $\alpha$. The total number of
displacement pulse combinations is thus quite large and
increases with the photon number $N$ in the NOON state,
typically involving of order a few hundred pulses. The
displacement pulses can be expressed as
\begin{eqnarray}
\label{eq.disp}
D_{AB} ( \alpha_j , \beta_k ) & = & D_A ( \alpha_j ) \otimes D_B ( \beta_k ) \nonumber\\
& = & e^{\alpha_j b_A^+ - \alpha_j^* b_A} \otimes e^{\beta_k b_B^+ - \beta_k^* b_B} \, .
\end{eqnarray}

\subsection{Photon Populations}
For an initial joint-resonator density matrix
$\rho_\mathrm{initial}$, the displacement pulses shift the
density matrix to
\begin{equation}
\label{eq.rho}
   \rho = D_{AB} ( - \alpha_j , - \beta_k ) \rho_\mathrm{initial} D_{AB} ( \alpha_j , \beta_k ) \, .
\end{equation}
By bringing both qubits (initially in their $| g \rangle$
states) on resonance with the resonators, the joint number
states contained in two resonators, i.e., the diagonal elements
of $\rho$, can be read out, as each diagonal element swaps with
the qubits at a different rate \cite{s_hofheinz:2009}, resulting in a
distinct time-dependence for the probabilities $P_{gg}$,
$P_{ge}$, $P_{eg}$, and $P_{ee}$. These can be
numerically simulated using the device parameters from
Table~\ref{tab.para}.

\begin{figure*}[t]
\includegraphics[width=0.9\textwidth]{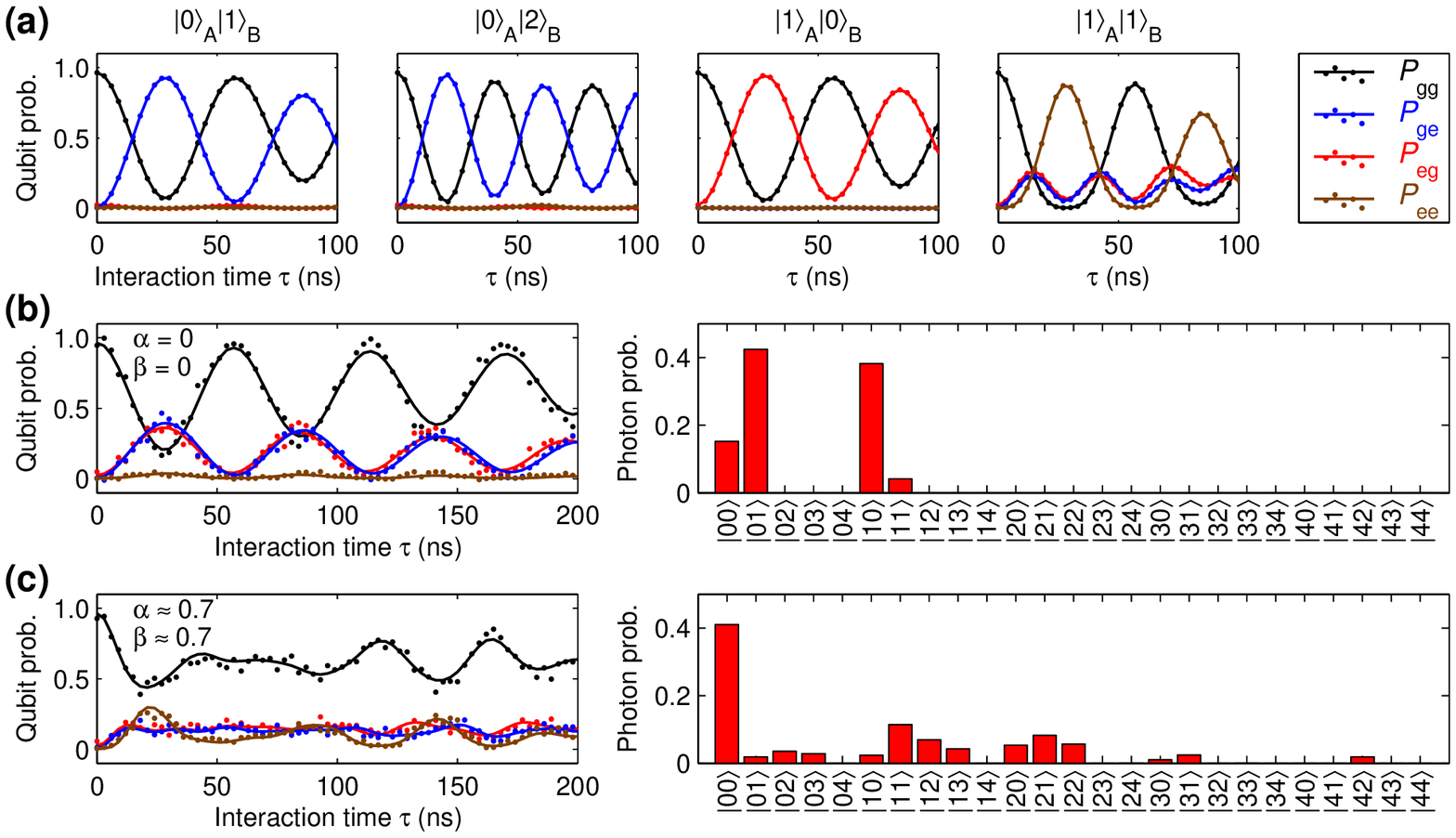}
\caption{\label{fig.PhotonNum} (Color online) Joint-state probabilities and
photon number distribution analysis for $N=1$ NOON
state. (a) Joint-qubit state
probabilities from numerical simulations, for a few selected
initial resonator states in $A$ and $B$, as indicated. Lines
are a guide to the eye. Device parameters are as in
Table~\ref{tab.para}. (b) Left panel, measured
joint-qubit state probabilities for null displacement pulses
$\alpha, \beta = 0$. Lines are fits to
the data. The fit joint-photon probability distributions
$\rho_{| m \rangle_A | n \rangle_B}$ are displayed in
the right panel. The most significant occupations are in $| 0
\rangle_A | 1 \rangle_B$ and $| 1 \rangle_A | 0
\rangle_B$, as expected for this state. (c) Left
panel, measured joint-qubit state probabilities for the
displaced state with $\alpha =0.7$ and
$\beta=0.7$ (square-root of photon number
units). Lines are fits to the data. The fit joint-photon
probability distributions $\rho_{ | m \rangle_A | n
\rangle_B }$, shown in the right panel, are significantly
different from those shown in (b) due to the
displacement. The matrix elements for the higher photon number
states are omitted for viewing, although these may not be
negligible for larger displacement pulses. }
\end{figure*}

As displayed in Fig.~\ref{fig.qubitTomo}(b), there is a small
non-zero occupation of the excited state of each qubit after
the state generation sequence, due to decoherence and pulse
imperfections. We use the measured qubit state after the state
generation sequence as the qubit initial condition when
numerically simulating the tomography. Using these simulations,
we obtain the time dependence for the each of the probabilities
$P_{gg}$, $P_{ge}$, $P_{eg}$, and $P_{ee}$
corresponding to different combinations of photon number (Fock)
states in the two storage resonators. Examples of these
probability traces are shown in Fig.~\ref{fig.PhotonNum}(a) for
some selected initial states. The time-dependent traces for
these probabilities, for the set of Fock states $\left \{
\left| m \right \rangle_A \left| n \right \rangle_B , m
= 0 , 1 , 2 , \ldots , n = 0 , 1 , 2 , \ldots
\right\}$, are then used to decompose the
experimentally-measured time traces, which yields the
probability distribution for the Fock number states contained
in the storage resonators. This thus yields the diagonal
elements of the experimentally-measured displaced density
matrix $\rho$.

We obtain the diagonal elements of $\rho$ by doing a
least-squares fit of the time-dependent probabilities, corrected
for measurement fidelity. We use the MATLAB packages YALMIP and
SeDuMi for the fitting. The number of fitting parameters is the
number of diagonal elements, directly determined by the maximum
photon number state contained in the resonators, plus the
number of photon quanta added by the displacement pulses $|
\alpha \rangle$ and $| \beta \rangle$. Fits are done with
constraints $P_{ m n } \geq 0$ and
$\sum_{m,n}{P_{ m n }} = 1$ to return
meaningful probability values. Examples of these fits are shown
in Fig.~\ref{fig.PhotonNum}, for the $N = 1$ NOON state.

\subsection{Joint-Resonator Density Matrices}

With the diagonal elements of $\rho$ measured for a set of
displacements $\left\{ D_{AB} ( \alpha_j , \beta_k )
\right\}$, $\rho_\mathrm{initial}$ can be solved for by
inverting Eq.~(\ref{eq.rho}) through a linear least-squares
fit, while restricting the density matrix to be Hermitian. The
resulting density matrices may have small negative eigenvalues
due to noise. We use the MATLAB packages YALMIP and SeDuMi to
perform semi-definite convex optimization programming, allowing
us to find a physical density matrix that is closest to the
actual matrix.

When solving for $\rho_\mathrm{initial}$ of a NOON state, we
restrict the dimension of $\rho_\mathrm{initial}$ to an $N
\times N$ matrix, even though the dimension of $\rho$ can be
significantly larger than this due to the displacement pulses.
We zero-pad the elements in $\rho_\mathrm{initial}$ that
have photon indices larger than $N$. This approach is validated
by the coincidence measurements (see main text), as we do not detect any frequency
components for number states above $N$ prior to injecting a
displacement pulse.

\section{NOON State Decay Dynamics}

\begin{figure*}[t]
\includegraphics[width=0.7\textwidth]{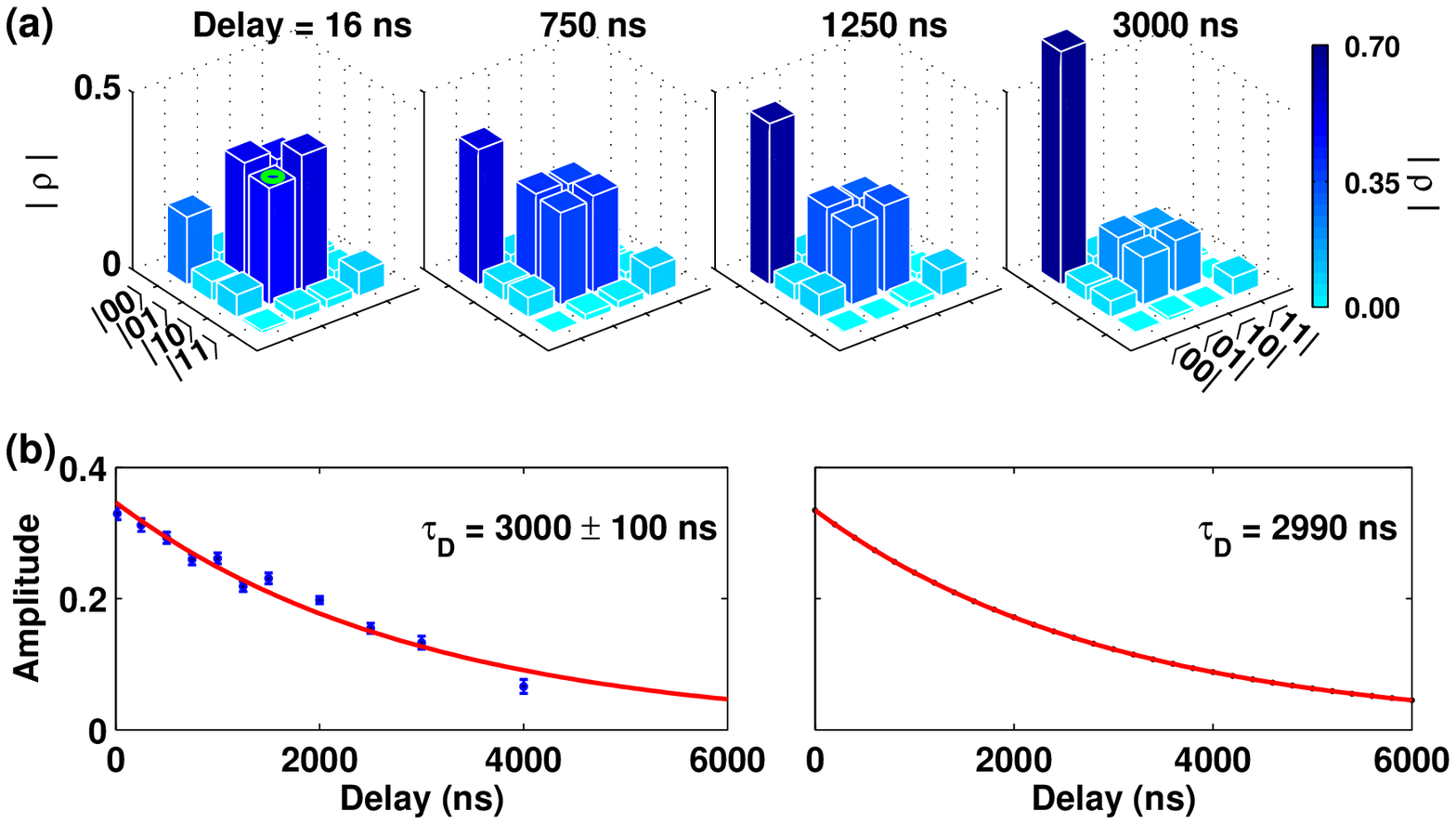}
\caption{\label{fig.decay} (Color online) Decay of the $N = 1$ NOON
state. (a) Density matrix amplitudes for the $N =
1$ NOON state at selected delay times. Bar heights and colors
represent the matrix element amplitude, scale on right. The off-diagonal element amplitudes decrease at a rate
close to that of the diagonal counterparts. Values for the
relevant elements are tabulated in Table \ref{tab.rho}. (b) Left: Experimental data, showing amplitude of the off-diagonal element as
marked by a green ellipse in the first panel in (a),
plotted versus post-preparation time. Line is a fit yielding a
single decay time $\tau_{D}$, consistent with a
Markovian dissipative environment \cite{s_walls:1985}. Right: Numerical simulation of the same density matrix element as the experiment, with the line an exponential fit. The simulation uses the
measured density matrix as the initial condition
and the resonator $T_1$ value listed in
Table \ref{tab.para}. The simulated decay time $\tau_D$ is in good agreement with experiment.}
\end{figure*}

The Wigner tomography allows us to study the decoherence dynamics
of the bipartite system \cite{s_wang:2009b}. The experimental
results, compared with numerical simulations, are shown in Fig. \ref{fig.decay}, with relevant elements in Table \ref{tab.rho}. We note that the
time evolution of the off-diagonal elements in the
two-resonator density matrix, which represent inter-resonator
coherence, is different from the evolution of the corresponding
off-diagonal elements for a single resonator, which represent
intra-resonator coherence \cite{s_wang:2009b}.

\begin{table}
\begin{tabular}{|c|c|c|c|c|} \hline \hline
Delay & & & & \\
(ns) & $\langle 0 1 | \rho | 0 1 \rangle$ & $\langle 1 0 | \rho | 1 0 \rangle$ & $\langle 0 1 | \rho | 1 0 \rangle$ & $\Delta | \langle 0 1 | \rho | 1 0 \rangle |$  \\\hline
16 & 0.36(1) & 0.386(9) & 0.328-0.037i & 0.009 \\
250 & 0.31(1) & 0.32(1) & 0.31+0.01i & 0.01 \\
500 & 0.32(1) & 0.31(1) & 0.285-0.068i & 0.009 \\
750 & 0.27(1) & 0.27(1) & 0.259+0.017i & 0.008 \\
1000 & 0.27(1) & 0.28(1) & 0.261+0.010i & 0.008 \\
1250 & 0.23(1) & 0.24(1) & 0.215-0.037i & 0.008 \\
1500 & 0.24(1) & 0.23(1) & 0.229-0.033i & 0.008 \\
2000 & 0.20(1) & 0.20(1) & 0.198+0.001i & 0.006 \\
2500 & 0.17(1) & 0.17(2) & 0.155-0.005i & 0.008 \\
3000 & 0.14(2) & 0.15(2) & 0.13+0.04i & 0.01 \\
4000 & 0.08(2) & 0.09(4) & 0.06+0.03i & 0.01 \\
\hline \hline
\end{tabular}
\caption{\label{tab.rho} Density matrix elements as a function
of delay time for the $N = 1$ NOON state shown in Fig. \ref{fig.decay}. Uncertainties for the diagonal elements are
in parentheses, while the magnitude of the uncertainty for the
off-diagonal terms are given in the last column.}
\end{table}

\subsection{Error Analysis}

Statistical errors in the qubit probability measurements as well as
uncertainty in the amplitude calibration for the displacement pulses $\alpha_j$ and $\beta_k$ are used to estimate the
uncertainty in the amplitude and phase of each density
matrix element. These errors are found to be small, in part because the constraints on the analysis filters unrealistic
values. Instead we find that slow phase drifts in the electronics,
perhaps dominated by ambient temperature fluctuations, give the main phase uncertainties, especially during
long measurements. Evaluating a single density matrix usually takes a relatively short time during which these drifts are minimal. However, measuring a
series of density matrices such as Fig. 5 in the main paper, takes a much longer time, typically $12$ to $20$ hours, allowing for more significant drifts. These mostly affect the phases of the density matrix elements, rather than the amplitudes.

\subsection{Validation}

\begin{figure*}[t]
\includegraphics[width=0.7\textwidth]{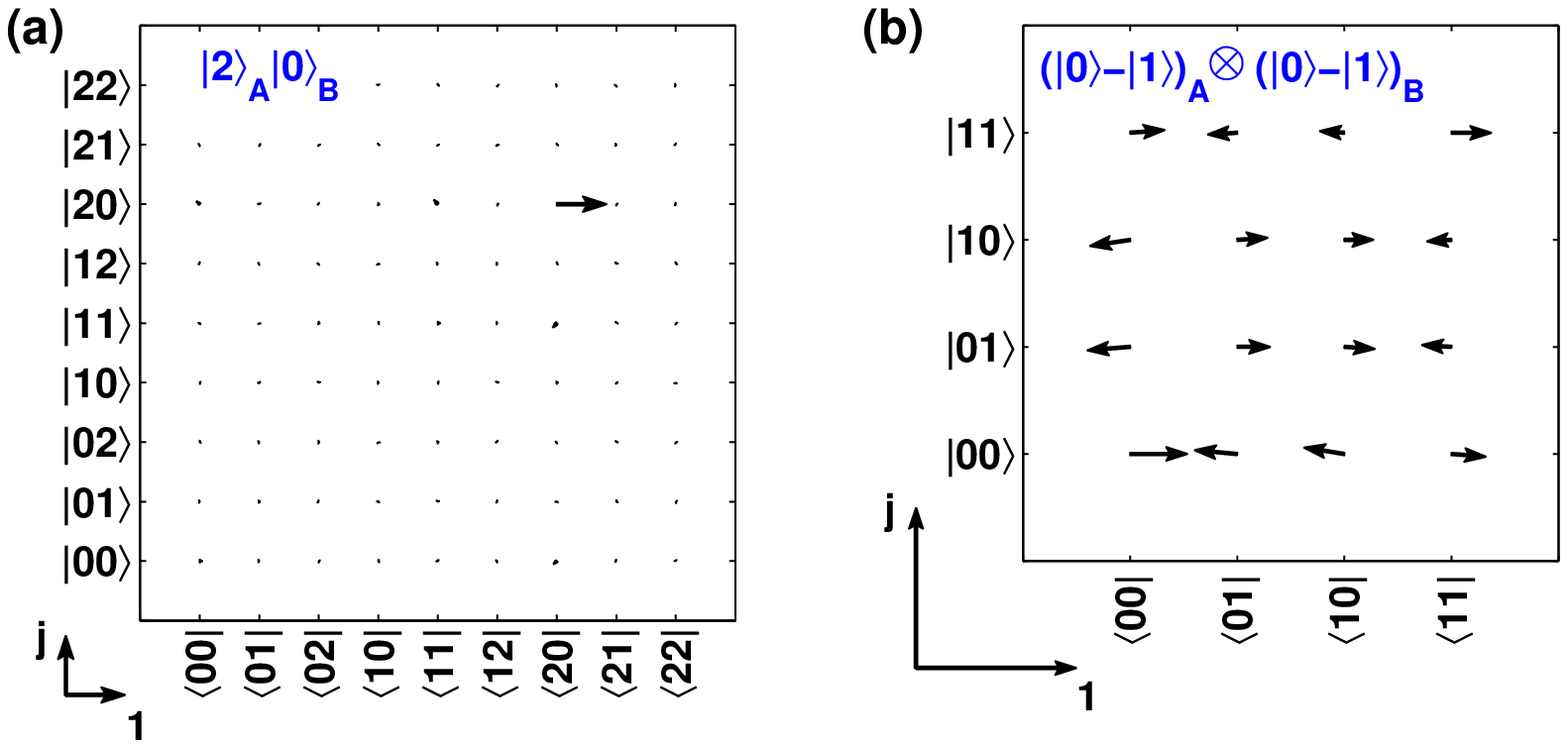}
\caption{\label{fig.extra}(Color online) Benchmark resonator states used to validate the
bipartite Wigner tomography. (a) Photon-number representation of the
two-resonator density matrix (error bars not shown), for the
energy eigenstate $| 2 \rangle_A | 0 \rangle_B$,
generated by pumping two photons into resonator $A$ while
leaving $B$ in its ground state. The magnitude of each matrix
element is represented by the full length of the corresponding arrow and the
phase angle determined by the direction of the arrow in the
complex plane (scale on bottom left). Density matrix is as
expected from state preparation. (b) Two-resonator
density matrix (error bars not shown) for the test resonator state
$( | 0 \rangle - | 1 \rangle )_A ( | 0 \rangle - | 1 \rangle
)_B$, generated by creating the $| g \rangle - i | e
\rangle$ state in each qubit, followed by a complete $i$-SWAP
transfer to each storage resonator. Non-idealities are likely
due to inaccuracies in the preparation pulses. }
\end{figure*}

The bipartite Wigner tomography was validated by several
consistency checks. (1) The density matrix is as expected for a range of different states,
including the highly entangled NOON states, the energy
eigenstates $| 2 \rangle_A | 0 \rangle_B$
(Fig.~\ref{fig.extra}(a)), the separable (product) state $( | 0 \rangle - |
1 \rangle )_A ( | 0 \rangle - | 1 \rangle )_B$
(Fig.~\ref{fig.extra}(b)) and the un-entangled mixed states (see below). (2) The NOON
states display the expected phase sensitivity as a function of
photon number $N$, as shown in Fig. 5 in the main paper and
Table \ref{tab.pha}. (3) The NOON state fidelity and
entanglement of formation agree reasonably well with numerical
simulations. (4) The time-dependence of the density matrix
elements, showing uniform exponential decay of all elements, is
as expected and agrees with numerical simulations, as shown in Fig. \ref{fig.decay} and Table \ref{tab.rho}. (5)
The calculated negativities are significantly above zero for
the NOON states and precisely zero (within the measurement
error) for the unentangled mixed state (see below).

\begin{table}
\begin{tabular}{|ccccc|} \hline \hline
& color & $\langle 0 1 | \rho | 0 1 \rangle$ & $\langle 1 0 | \rho | 1 0 \rangle$ & $\langle 0 1 | \rho | 1 0 \rangle$ \\ \hline \hline
\multirow{3}{*}{Fig.~5(a)}& black & 0.327 & 0.460 &  -0.229+0.239i \\
& blue & 0.318 & 0.464 & 0.204+0.200i \\
& red & 0.301 & 0.439 & 0.196-0.156i \\\hline\hline
\comments{\multirow{3}{*}{Fig.~3C ($N = 2$)}& black & 0.280 & 0.370 & -0.033-0.186i \\
& blue & 0.283 & 0.370 & 0.124-0.145i \\
& red & 0.266 & 0.366 & 0.173-0.052i \\ \hline}
\end{tabular}
\caption{\label{tab.pha} (Color online) Density matrix elements for the data
shown in Figs.~5(a) of the main paper, coded by color
(uncertainties not shown). The phase uncertainty in the
off-diagonal elements is dominated by the slow phase drift of
our electronics (see text).}
\end{table}

We note that bipartite Wigner tomography can measure any
matrix element with a relatively high accuracy, as we can
displace the system by an arbitrary amount in phase space. Even for relatively small
off-diagonal elements, such as the desired off-diagonal term in
the $N = 3$ NOON state, the tomography can unambiguously evaluate this element and measure its sensitivity to
external phase perturbations.

\subsection{Ensemble of Mixed States}

We use a synthetic ensemble of mixed states to illustrate the
hazards involved in relying purely on coincidence measurements
for demonstrating NOON-state entanglement. The ensemble comprises a mixture of 50\% $| 1 \rangle_A | 0 \rangle_B$ and 50\% $| 0 \rangle_A | 1 \rangle_B$ states, i.e., an ensemble with the same probability of being measured in either of the states forming the $N=1$ NOON state, but without any entanglement. This is done by generating the pure state $| 1 \rangle_A | 0 \rangle_B$ and measuring the time-dependent joint probabilities $P_{gg}(\tau)$,
$P_{ge}(\tau)$, $P_{eg}(\tau)$, and $P_{ee}(\tau)$ for this state. We then
generate the other component of the ensemble, the pure state $| 0 \rangle_A | 1 \rangle_B$, and
repeat the generation and measurement procedure. Each value of $\tau$ involves 300 repeats of the preparation and measurement sequence for each of the pure states. We then
combine the measurement results with equal weights, creating the joint
probabilities for the synthetic ensemble; these
data are shown in the main paper. The tomographic analysis yielding the density matrices is
done in the usual way. The outcome of
the ensemble measurements are shown in the main text, with the
joint probabilities evolving in a way indistinguishable from
the entangled NOON states, but the density matrix for the
ensemble revealing a complete lack of entanglement, as
witnessed by the negligible values for the off-diagonal
elements.

\section{MOON State}

\begin{figure}
\includegraphics[clip=True,width=0.48\textwidth]{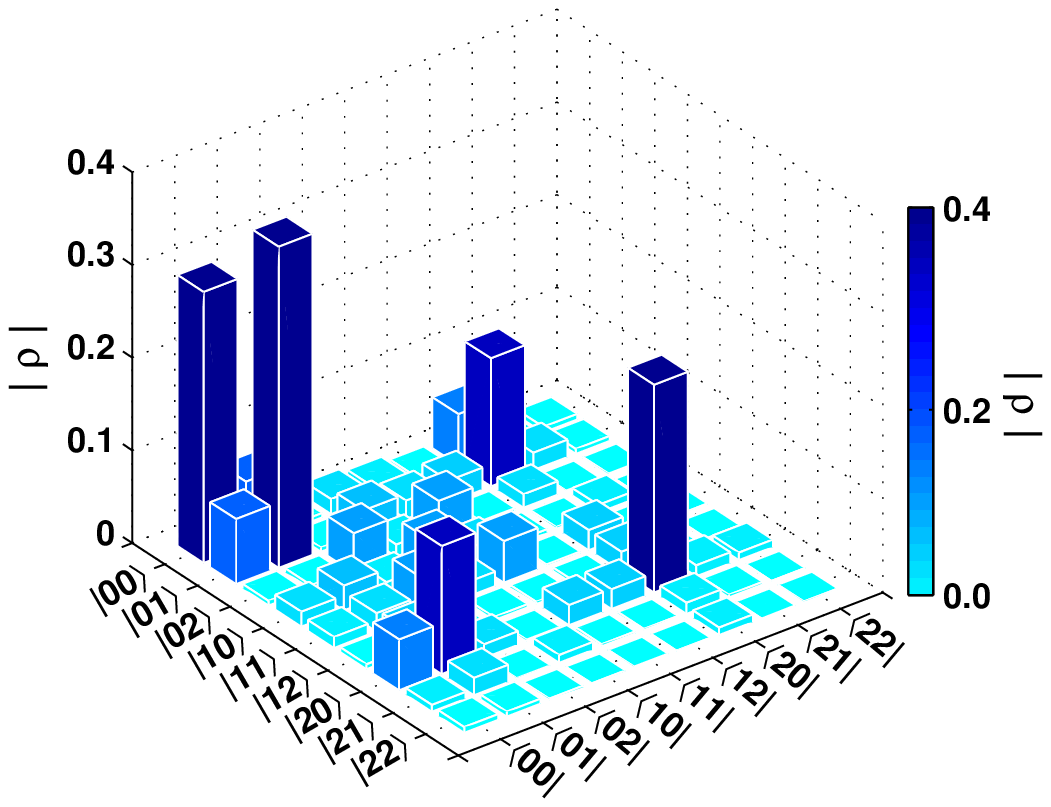}
\caption{\label{fig.moon} (Color online) Measured density matrix (errors not
shown) of the MOON state with $M = 2$ and $N = 1$.
The generation sequence is as described in the text. The state
fidelity is $0.42 \pm 0.01$, lower than our typical NOON state
fidelity, due to technical issues with the particular device
used for this experiment; the entanglement of formation is 0.16. The negativity is $N_e = 0.14 \pm 0.01$,
indicating a statistically significant entanglement.}
\end{figure}

The NOON-state generation protocol can be simply generalized to
generate MOON states, with different photon numbers in the two
entangled resonators. The generation is similar to the NOON
state sequence shown in Fig. \ref{fig.qubitTomo}. We assume $M
> N$: After generating the Bell entanglement between two qubits
at the end of step I (Fig. \ref{fig.qubitTomo}(a)), we repeat
step II $N-1$ times, yielding the four-fold entangled state $|
e g ( N - 1 ) 0 \rangle + | g e 0 ( N - 1 ) \rangle$. The
photon amplification and transfer process (step II) is then
applied $M-N$ times, but only to qubit $q_0$ and resonator
$A$, yielding the state $| e g ( M - 1 ) 0 \rangle + | g e 0 (
N - 1 ) \rangle$. The final qubit excitations are then
transferred in step III, resulting in the MOON state $| g g
\rangle \otimes \left ( | M 0 \rangle + | 0 N \rangle \right
)$, with the qubits disentangled from the resonators. A MOON state generated in this fashion, with $M = 2$
and $N = 1$, is shown in Fig.~\ref{fig.moon}.

\end{document}